\documentclass[12pt,twoside]{article}

\usepackage[usenames]{color}
\usepackage{lscape}
\usepackage{amssymb}
\usepackage{rotating}
\usepackage[T1]{fontenc}
\usepackage{ae}
\usepackage{aecompl}
\usepackage{pifont}
\usepackage{graphicx}
\usepackage{fancyhdr}
\usepackage{amsmath}
\usepackage{times}

\def\pa{\partial}
\def\nn{\nonumber \\}
\def\ov{\overline}
\def\br{&\!\!\!\!}

\newlength{\dinwidth}
\newlength{\dinmargin}
\begin{document}

\thispagestyle{empty}
\begin{flushright}
\end{flushright}

\vspace*{15mm}

\begin{center}
{\Large\bf
Volume modulus inflation
and a low scale of SUSY breaking
}
\vspace*{5mm}
\end{center}
\vspace*{5mm} \noindent
\vskip 0.5cm
\centerline{\bf
M. Badziak and M. Olechowski
}
\vskip 5mm
\centerline{\em Institute of Theoretical Physics,
University of Warsaw}
\centerline{\em ul.\ Ho\.za 69, PL--00--681 Warsaw, Poland}

\vskip 15mm

\centerline{\bf Abstract}
\vskip 3mm

The relation between the Hubble constant and the scale of supersymmetry
breaking is investigated in models of inflation dominated by a
string modulus. Usually in this kind of models the gravitino
mass is of the same order of magnitude as the Hubble constant which
is not desirable from the phenomenological point of view.
It is shown that slow-roll saddle point inflation may be
compatible with a low scale of supersymmetry breaking only
if some corrections to the lowest order K\"ahler potential
are taken into account. However, choosing an appropriate  K\"ahler
potential is not enough. There are also conditions for the
superpotential, and e.g.\ the popular racetrack
superpotential turns out to be not suitable.
A model is proposed in which slow-roll inflation and
a light gravitino are compatible. It is based on a superpotential
with a triple gaugino condensation and the K\"ahler potential with
the leading string corrections. The problem of fine tuning
and experimental constraints are discussed for that model.

\newpage

\section{Introduction}

The existence of an inflationary stage in the very early Universe
is a paradigm of the contemporary cosmology.
Inflation solves many problems of the standard cosmology, such as
flatness and isotropy of the observed Universe. It provides also the
best known mechanism to generate the primordial density fluctuations.
This feature of inflation makes it testable by means of the CMB
spectrum which is measured very precisely by WMAP and will be measured
even better by the forthcoming Planck satellite.

Inflation is usually implemented via dynamics of a scalar field -
the inflaton. The scalar sector of the Standard Model contains only
the Higgs field and its potential cannot accommodate inflation.
Thus, inflation can be realized only in some (more fundamental)
generalization of the Standard Model.
The most promising candidates for ``the theory of everything''
are 10-dimensional string theories. To make contact with
our low-energy 4-dimensional world, six of these dimensions have
to be compactified.
The main obstacle which for many years prevented from doing
phenomenology within string theories was lack of a potential for
the moduli fields parametrizing 6-dimensional internal manifolds.
The breakthrough in the moduli stabilization was made
within the framework of type IIB string theory, the dilaton and the
complex structure moduli (CSM) were stabilized by turning on
some non-trivial fluxes \cite{Giddings}.
A mechanism to stabilize also the K\"ahler
moduli, including the volume modulus, was proposed in the famous KKLT
model \cite{kklt}. It uses nonperturbative effects, such as the gaugino
condensation, which give rise to terms in the
superpotential\footnote{
In the KKLT model it is assumed that the dilaton and the CSM are
  stabilized at high energies and nonperturbative effects do not destabilize
  them. In \cite{stab} a detailed analysis of the validity of this assumption
  was done. It was found that in the simplest model without CSM moduli
  stabilization cannot be achieved.
}
depending exponentially on the volume modulus. As a result, the volume
modulus is stabilized in a supersymmetric (SUSY) anti de Sitter (AdS)
minimum which, after inclusion of anti-branes, is uplifted to a
de Sitter (dS) space.
The main drawback of this model is the explicit breaking of SUSY
by anti-branes. However, this part of the KKLT model has been improved
and the moduli have been stabilized in dS vacua with SUSY broken
spontaneously by F-terms \cite{fuplifting}-\cite{achucarro2}
or D-terms \cite{duplifting}-\cite{dudas}.

Development of dS string vacua opened the possibility of constructing
inflationary models within string theories. There are two types of
scenarios: One is brane inflation, where the interbrane distance
plays the role of the inflaton \cite{braneinf}-\cite{braneinf2}.
The other one is
moduli inflation, where the inflaton is one of the moduli fields.
In this paper, we concentrate on the latter scenario. In the KKLT
model with only one exponential term in the superpotential, the
potential is too steep for slow-roll inflation. However, adding
second exponential term to the superpotential makes inflation possible.
This was done in \cite{racetrack} where a model called the racetrack
inflation was proposed.
It is a model with one K\"ahler modulus, the volume modulus.
Its imaginary part plays the role of the inflaton.
Similarly as in the KKLT model, a dS vacuum is obtained
by non-supersymmetric uplifting. Racetrack inflation models with
supersymmetric uplifting have been also constructed.
A model with string theory $\alpha'$-corrections as a source of
uplifting and with SUSY broken in a dS vacuum by nonvanishing F-terms
was presented in \cite{westphal}.
D-terms were used to uplift the potential in racetrack inflation
in \cite{brax}.
The racetrack inflation model was generalized to the case of
two K\"ahler moduli in \cite{betrace}.
Other models of moduli inflation were
proposed in \cite{lalak2}-\cite{conlon}.

It was pointed out in \cite{kl} that in inflationary models based on
the KKLT moduli stabilization, the gravitino mass is typically of the
order of the Hubble scale during inflation which
should be many orders of magnitude larger than the electroweak scale.
Models with such heavy gravitino (typically much above the TeV scale)
are disfavored from the phenomenological point of view.
A possible solution to this problem was proposed in \cite{kl}
where it was observed that in models with a SUSY Minkowski
vacuum the gravitino mass is not directly related to the scale of inflation.
Such SUSY Minkowski vacua exist in KKLT type models
with the racetrack superpotentials (an additional tuning of
parameters is necessary). We will call it the Kallosh-Linde (KL) model.

In \cite{accinf} an inflationary model based on the KL model was
constructed. However, in this model the moduli are stabilized in a
non-SUSY Minkowski minimum, which was obtained by uplifting an AdS
minimum (existing in addition to the SUSY Minkowski one).
Therefore, in this model the gravitino mass is also much larger than
the TeV scale.

The main goal of this paper is to construct an inflationary model,
within the framework of type IIB string theory, with the gravitino
mass much smaller than the Hubble constant during inflation.
We restrict ourselves to models with only one K\"ahler modulus
(the volume modulus) and assume that the dilaton and the CSM are
stabilized by fluxes at some higher scales. The results of our
investigation should be valid also for multi-field models with
inflation dominated by the volume modulus. We focus on inflation
that occurs in the vicinity of the saddle point with the inflaton
rolling down towards the SUSY  Minkowski minimum at which the
gravitino mass vanishes (or near-Minkowski minimum at which the
gravitino mass is very small). In order to find flat enough saddle
points, which are necessary for slow-roll inflation, we perform a
general study of non-SUSY stationary points with arbitrary K\"ahler
potential and superpotential. We derive a necessary condition for
slow-roll inflation. Then, we focus on the string inspired K\"ahler
potentials and show that, for a tree-level K\"ahler potential and
an arbitrary superpotential, inflation that finishes in the SUSY
Minkowski minimum cannot be realized. We find that the perturbative
corrections to the K\"ahler potential can improve somewhat the
situation but even with such corrections, inflation still cannot
be implemented in the KL model.

We propose a model in which inflation with the Hubble constant much
bigger than the gravitino mass can be realized. It is based on a
superpotential with three exponential terms that may originate from
the gaugino condensation in a hidden sector. We use also the K\"ahler
potential with $\alpha'$-corrections and string loop corrections.
In this setup successful slow-roll inflation can be obtained with
the spectral index consistent with the observations. The inflaton,
which is mainly the imaginary part of the volume modulus, rolls
down towards the SUSY near-Minkowski minimum, where inflation ends.
Therefore, in this model the gravitino mass can be made very small.

The paper is organized as follows.
In section 2 we analyze non-SUSY stationary points
and formulate conditions necessary for slow-roll inflation.
In section 3 we show how the string inspired corrections to
the K\"ahler potential may help in fulfilling such conditions.
The KL model is analyzed in section 4. We show that slow-roll
inflation can not be realized in racetrack models with
SUSY Minkowski minimum even with the corrected K\"ahler potential.
In section 5 we propose a triple gaugino condensation model.
It can accommodate slow-roll inflation with the
Hubble constant much bigger than the gravitino mass.
We study predictions of this model and show that they
are compatible with current observational status. Finally,
we conclude in section 6.

\section{Non-supersymmetric stationary points}

The scalar potential in supergravity can be expressed in terms of
the superpotential $W$ and the K\"ahler potential $K$ in the
following way\footnote{
We use Planck units where $M_p=1$.
}:
\begin{eqnarray}
        \label{potentialWK}
        V=e^K\left(K^{I\ov{J}}D_I W\ov{D_J W}-3\left|W\right|^2\right) \ .
\end{eqnarray}
Supersymmetric stationary points of this potential satisfy the
condition:
\begin{eqnarray}
        \label{susycond}
        D_I W=\pa_I W+\pa_I KW=0 \ .
\end{eqnarray}
Using (\ref{susycond}) and (\ref{potentialWK}), we immediately see
that the value of the potential at a SUSY stationary point is
always non-positive and vanishes only when
\begin{eqnarray}
        \label{minkcond}
        \pa_I W=0\ ,\hspace{4cm} W=0 \ .
\end{eqnarray}
Models with SUSY Minkowski vacua within type IIB string theory
were studied in \cite{krefl}. In \cite{mink} it was shown that any
Minkowski vacuum, satisfying conditions (\ref{minkcond}), is stable.
Of course, the gravitino mass vanishes in a SUSY Minkowski vacuum.

We are interested in inflation ending in a SUSY (near) Minkowski vacuum.
Inflation may end in a Minkowski vacuum if it starts
from the vicinity of a (nearby) saddle point with positive
energy. In \cite{scrucca} the necessary conditions
for the stability of non-SUSY Minkowski vacua were found.
In what follows, we generalize those results for any non-SUSY
stationary points.

For this analysis, it is convenient to work with function $G$
defined by:
\begin{eqnarray}
        \label{G}
        G(\Phi_I,\Phi_I^{\dag})=K(\Phi_I,\Phi_I^{\dag})
+\log W(\Phi_I)+\log \ov{W}(\Phi_I^{\dag}) \ .
\end{eqnarray}
In terms of $G$, the scalar potential can be written
as\footnote{
We use the standard notation $G_I\equiv\frac{\pa G}{\pa\Phi_I}$,
$G_{\ov{I}}\equiv\frac{\pa G}{\pa\Phi_I^\dag}$.
}:
\begin{eqnarray}
        \label{potentialG}
        V=e^G\left(G^{I\ov{J}}G_I G_{\ov{J}}-3\right) \ .
\end{eqnarray}
Following \cite{scrucca}, we use the tools of K\"ahler geometry with
the metric given by the second derivative of the K\"ahler
potential $G_{I\ov{J}}$.
A covariant derivative of a scalar is equal to an ordinary
derivative, therefore we can write the stationarity conditions
using covariant derivatives (which is more convenient):
\begin{eqnarray}
        \label{stationarity}
        G_I(G^K G_K -2)+G^K\nabla_I G_K=0 \ .
\end{eqnarray}
The second covariant derivatives of the potential are given by
\begin{eqnarray}
\nabla_I\nabla_J V=\pa_I\pa_J V-\Gamma^K_{IJ}\pa_K V \ ,
\label{nablanablaV}
\end{eqnarray}
where $\Gamma^K_{IJ}$ is the connection for the K\"ahler manifold
defined by the metric $G_{I\ov{J}}$. We are interested in the second
derivatives at stationary points, where the first derivatives vanish.
At such points, the term in (\ref{nablanablaV}) proportional to the
connection vanishes, and the ordinary second derivatives
are equal to the covariant second derivatives.
Therefore, the matrix of the second derivatives of the potential
at a stationary point reads:
\begin{eqnarray}
\begin{pmatrix}
\partial_{I}\partial_{\ov{J}}V & \partial_I\partial_{{J}}V  \\
\partial_{\ov{I}}\partial_{\ov{J}}V & \partial_{\ov{I}}\partial_{J}V
        \end{pmatrix}
=
\begin{pmatrix}
                V_{I\ov{J}} & V_{IJ} \\
        V_{\ov{I}\ov{J}} & V_{\ov{I}J}
        \end{pmatrix} \ ,
        \end{eqnarray}
where the second covariant derivatives,
$V_{I\ov{J}}=\nabla_I\nabla_{\ov{J}}V$ and $V_{IJ}=\nabla_I\nabla_J V$,
are given by the following expressions:
\begin{eqnarray}
        \label{2derV}
V_{I\ov{J}}\br=\br e^G\left(G_{I\ov{J}}(\widehat{G}^2-2)
-G_I G_{\ov{J}}(\widehat{G}^2-3)
+\nabla_I G_K\nabla_{\ov{J}}G^K-R_{I\ov{J}K\ov{L}}G^KG^{\ov{L}}\right)
\ , \\[6pt]
        \label{2derV2}
V_{IJ}\br=\br e^G\left((\nabla_I G_J+\nabla_J G_I)\frac{\widehat{G}^2-1}{2}
-G_I G_J(\widehat{G}^2-3)+\frac{1}{2}G^K
\left\{\nabla_I,\nabla_J\right\}G_K\right)
\ ,
\end{eqnarray}
$V_{\ov{I}J}=\ov{V_{I\ov{J}}}$, $V_{\ov{I}\ov{J}}=\ov{V_{IJ}}$ and all
quantities should be understood as evaluated at a stationary point.
We introduced a quantity $\widehat{G}\equiv\sqrt{G^I G_I}$ which is
related in a simple way to the value of the potential:
\begin{eqnarray}
\label{gv}
\widehat{G}^2=3+e^{-G}V \ .
\end{eqnarray}
For $\widehat{G}^2=3$, which corresponds to the Minkowski condition,
in each of the eqs.\ (\ref{2derV}) and (\ref{2derV2}) the first term
simplifies while the second one vanishes.
In case of dS stationary points, which are of main interest
in this paper, we have $\widehat{G}^2>3$.

Expressions (\ref{2derV}) and (\ref{2derV2}) were derived in full
generality but to use them in practice one has to impose some restrictions.
For the purpose of this work, it is enough to restrict
to the one-field case. For non-canonically normalized fields
the physical mass matrix is given by
\begin{eqnarray}
M^2=\begin{pmatrix}
                m_{X\ov{X}}^2 & m_{XX}^2 \\[4pt]
        m_{\ov{X}\ov{X}}^2 & m_{\ov{X}X}^2
        \end{pmatrix} \ ,
\label{M2}
        \end{eqnarray}
with the entries
\begin{eqnarray}
m_{X\ov{X}}^2=\frac{V_{X\ov{X}}}{G_{X\ov{X}}}
\ ,\hspace{1cm}
m_{XX}^2=\frac{V_{XX}}{G_{X\ov{X}}}\ ,
\end{eqnarray}
which can be written in the following form:
\begin{eqnarray}
\label{mxxb}
m_{X\ov{X}}^2\br=\br e^G\left(2-{\widehat{G}}^2R_X\right) ,\\[6pt]
\label{mxx}
m_{XX}^2\br=\br\theta_X^2e^G\left[-2\left(\widehat{G}^4-3\widehat{G}^2+1\right)
+\widehat{G}^4A_{XXX}+3\widehat{G}^4\left(\widehat{G}^2-2\right)A_{XX\ov{X}}
-\widehat{G}^6A_{XXX\ov{X}}\right],
\end{eqnarray}
where
$\theta_X^2\equiv G_X/G_{\ov{X}}$, $A_{XX}\equiv G_{XX}/G_XG_X$,
$A_{X\ov{X}}\equiv G_{X\ov{X}}/G_XG_{\ov{X}}$, etc\ldots.
The curvature scalar of the K\"ahler manifold, $R_X$, is given by:
\begin{eqnarray}
R_X=\frac{G_{XX\ov{X}\ov{X}}}{G_{X\ov{X}}^2}
-\frac{G_{XX\ov{X}}G_{X\ov{X}\ov{X}}}{G_{X\ov{X}}^3}
\,.
\end{eqnarray}
Notice that all quantities in $R_X$ contain derivatives of $G$ with respect
to both holomorphic and anti-holomorphic variables. So, $R_X$
does not depend on the superpotential (which is holomorphic)
and is fully determined by the K\"ahler potential $K$.
In the case of stationary points
satisfying the Minkowski condition, i.e. $\widehat{G}^2=3$, the diagonal entry
$m_{X\ov{X}}^2$ of the mass matrix (\ref{M2}) depends on the superpotential
$W$ only via the overall factor $e^G$ because the Minkowski condition
fixes the value of $\widehat{G}^2$. On the other hand, looking for a saddle
point appropriate for inflation, we do not insist on any
particular value of the potential at such point. It only has to
be positive.
Therefore, we do not fix the value of $\widehat{G}^2$, and from (\ref{mxxb})
one can see that $m_{X\ov{X}}^2$
does depend on both the K\"ahler potential and the superpotential.

In the one-field case the mass eigenvalues can be computed analytically:
\begin{eqnarray}
m_{\pm}^2=m_{X\ov{X}}^2\pm \left|m_{XX}^2\right| \ .
\end{eqnarray}
When constructing a model of inflation, one usually deals with real fields
rather than with complex ones. It is convenient to introduce a new object,
the so-called $\eta$-matrix, which is defined for real fields and
is very useful when looking for models appropriate for inflation.
The entries of the $\eta$-matrix are given by the second
covariant derivatives with respect to real fields in the following way:
\begin{eqnarray}
\eta_i^j=\frac{g^{jk}\nabla_i\nabla_kV}{V} \ ,
\end{eqnarray}
where $g^{jk}=G^{J\ov{K}}/2$ and the lower case indices
correspond to imaginary or real parts of complex fields
(represented by capital letter indices).
The smallest eigenvalue of the $\eta$-matrix is a multi-field
generalization of the slow-roll parameter $\eta$. Inflation can take
place in the vicinity of a saddle point for which the parameter $\eta$
is very small and negative while all other eigenvalues of the $\eta$-matrix
are positive. In other words, this saddle point should be very flat
in the unstable direction.
At the stationary points the entries of the $\eta$-matrix are
proportional to the corresponding entries of the mass-matrix, and
this flatness condition, in the one-field case, can be formulated as:
\begin{eqnarray}
0<\left|m_{XX}^2\right|-m_{X\ov{X}}^2\ll V_0 \,,
\end{eqnarray}
where $V_0$ is the value of the potential at the stationary point.
In principle, one could use the above condition together with (\ref{mxxb})
and (\ref{mxx}) to look for models suitable for inflation.
However, $m_{XX}^2$ depends on $K$ and $W$ in a very complicated
way (eq.\ (\ref{mxx})), so it may turn out to be a highly non-trivial task.
On the other hand, a necessary condition for a successful model
of inflation says that the trace of the $\eta$-matrix is
positive\footnote{
To be strict, slow-roll inflation could be possible also for
slightly negative trace of the $\eta$-matrix. It would require one
very small negative eigenvalue and the other one with even
smaller absolute value.
However, this would require more fine tuning of parameters.
It is also very unlikely from the observational point
of view because it would provide very significant production of
isocurvature fluctuations, which has not been observed.
For the recent study on the isocurvature perturbations see e.g. \cite{iso}.
}.
It is relatively simple:
\begin{eqnarray}
\label{neccond}
m_{X\ov{X}}^2>0 \ ,
\end{eqnarray}
and, using eq.\ (\ref{mxxb}), can be rewritten as a condition
for the K\"ahler curvature:
\begin{eqnarray}
\label{rxcond}
R_X<\frac{2}{\widehat{G}^2} \ .
\end{eqnarray}
This condition involves both the K\"ahler potential and the
superpotential. However, a superpotential-independent upper bound
on the value of the K\"ahler curvature can be found for all
dS stationary points (which are most interesting for inflation).
It follows from (\ref{gv}) and (\ref{rxcond}) that the parameter
$\eta$ may be small only for such stationary points for which
\begin{eqnarray}
\label{rxncond}
R_X<\frac{2}{3} \ .
\end{eqnarray}
This condition is weaker than (\ref{rxcond}) but it is still
an important one because it can be used to eliminate
some models even without specifying the superpotential.
It should be stressed that (\ref{rxncond}) is necessary but not
sufficient to satisfy the condition (\ref{neccond}).
Notice also that the right-hand side of (\ref{rxcond}) is always positive,
so (\ref{neccond}) is satisfied when $R_X$ is negative or zero.

\section{String inspired K\"ahler potentials}

We concentrate now on the class of models motivated by string theories
for which the K\"ahler potential is given by:
\begin{eqnarray}
K=-n_X\ln(X+\ov{X}) \ ,
\label{standardK}
\end{eqnarray}
where $n_X$ is a positive integer. For the above K\"ahler potential
the curvature scalar is constant and has a very simple form:
\begin{eqnarray}
\label{stringcon}
R_X=\frac{2}{n_X} \ .
\end{eqnarray}
Using (\ref{mxxb}) and (\ref{gv}) we can formulate the following
condition necessary for the existence of a flat saddle
point (\ref{neccond}):\footnote{
A similar result, in a different context, was obtained in \cite{nilles}.
}
\begin{eqnarray}
e^G(n_X-3)-V_0>0 \ .
\end{eqnarray}
Inflation has to start from a saddle point satisfying the above
condition with a positive energy.
It is clear that no such
saddle points exist in models with $n_X\leq3$.
Whether they exist or not for $n_X>3$, depends on details of
a specific model.

For the K\"ahler potential (\ref{standardK}) we find the
following expression for the trace of the $\eta$-matrix:
\begin{eqnarray}
\eta_x^x+\eta_{\chi}^{\chi}
=-\frac{4}{n_X}\left(1-\frac{e^G\left(n_X-3\right)}{V_0}\right) \ ,
\label{tretaX}
\end{eqnarray}
where $x=$ Re$X$ and $\chi=$ Im$X$. The r.h.s.\ of this equation
is negative for any $V_0>0$ and $0<n_X\le3$. It is a significant
result which tells us that for any superpotential $W$ and the
standard K\"ahler potential, slow-roll inflation dominated by one
modulus and starting close to a saddle point of the potential is
not possible in a broad class of models inspired by string theories.

Equation (\ref{tretaX}) simplifies in type IIB string theory
(on which we focus in this paper) for which the K\"ahler potential
for the volume modulus is given by:
\begin{eqnarray}
\label{kahler}
K=-3\ln(T+\ov{T}) \ .
\end{eqnarray}
For this setup the trace of the $\eta$-matrix takes
a constant negative value:
\begin{eqnarray}
\label{treta}
\eta_t^t+\eta_{\tau}^{\tau}=-\frac{4}{3} \ ,
\end{eqnarray}
where $t=$ Re$T$ and $\tau=$ Im$T$.

The condition (\ref{treta}) forbids saddle point inflation in
supergravity theories with the K\"ahler potential (\ref{kahler}).
So, how is it possible that the original racetrack
model \cite{racetrack} successfully implements inflation, even
though it starts from the vicinity of a non-SUSY saddle point?
To answer this question, we recall that a key ingredient
of the racetrack model is the uplifting term in the potential
\begin{eqnarray}
\Delta V=\frac{E}{t^2} \ ,
\end{eqnarray}
which explicitly breaks supersymmetry. The numerical value of $E$
is fine tuned in order to uplift the AdS minimum to the Minkowski one.
It turns out that this term plays also a very important role from the
inflationary point of view. The reason is that the uplifting gives
an additional contribution to the second derivative of the potential
with respect to $t$ increasing $\eta_t^t$ by
\begin{eqnarray}
\Delta\eta_t^t
=
\frac{1}{V_0}\frac{4E}{t^2}+\ldots \ ,
\end{eqnarray}
where ellipsis denotes corrections due to different position of
the saddle point and the value of the potential after uplifting.
Other entries of the $\eta$-matrix are also changed.
Slow-roll inflation is possible when
$\Delta\eta_t^t+\Delta\eta_{\tau}^{\tau}$ is big enough at the
uplifted saddle point. We have checked that for the parameters used
in \cite{racetrack}, the value of
$\Delta\eta_t^t+\Delta\eta_{\tau}^{\tau}$ is about 4, which is
substantially bigger than the limiting value $\frac{4}{3}$.
Parameter $E$ has to be large enough to increase appropriately
the trace of the $\eta$-matrix. So, the SUSY AdS minimum before
uplifting has to be rather deep to ensure that after uplifting
$\Delta\eta_t^t+\Delta\eta_{\tau}^{\tau}$ is big enough.
As we mentioned before, a deep AdS minimum is the source of
a large gravitino mass in the uplifted Minkowski minimum.

There are also racetrack inflation models with uplifting which
breaks SUSY spontaneously by F-terms \cite{westphal}
or D-terms \cite{brax}. Such upliftings contribute to
the $\eta$-matrix in a more complicated way but in every
case they make the trace of the $\eta$-matrix positive.
The main disadvantage of this class of models is a large gravitino
mass. In the following sections we discuss mechanisms which
can change the condition (\ref{treta}) without increasing a small
(or even vanishing) gravitino mass.

\subsection{Corrections to K\"ahler potential}

In this subsection we identify corrections to the K\"ahler potential
which may change the sign of the $\eta$-matrix trace.
A necessary condition for the positivity of that trace is
\begin{equation}
R_T<\frac{2}{3}
\,,
\label{nX3}
\end{equation}
while a sufficient one is
\begin{equation}
\label{rt0}
R_T\leq0 \ .
\end{equation}
In a type IIB model with the leading order of the K\"ahler potential
for the volume modulus $T$ (\ref{kahler}), the K\"ahler curvature
equals $R_T=2/3$. Therefore, the necessary condition (\ref{nX3}) in
such a model is only marginally violated, so even relatively small
corrections to $K$ can make this condition satisfied. On the other
hand, $R_T=2/3$ is far away from the sufficient condition (\ref{rt0}).

In general, it is not hard to satisfy condition (\ref{rt0}).
For example, in the case of the simplest K\"ahler potential
$K=XX^{\dag}$, which provides the canonical kinetic term for
the field $X$, $R_X$ vanishes.
Unfortunately, this is not the case for the moduli fields.
We check now whether corrections to the leading order
K\"ahler potential (\ref{kahler}) can give a negative contribution
to $R_T$. We consider the corrections of the following form:
\begin{eqnarray}
\label{cork}
\Delta K=-\frac{\xi}{(T+\ov{T})^k} \ .
\end{eqnarray}
The scalar curvature for the corrected K\"ahler metric reads:
\begin{eqnarray}
R_T=\frac{2}{3}-\frac{k(k-1)(k+1)(k+2)}{9}\frac{\xi}{(T+\ov{T})^k}
+{\cal O}(\xi^2) \ .
\end{eqnarray}
As one can see, for $\xi>0$ and $k>1$, correction (\ref{cork})
gives a negative contribution to $R_T$.
The string theoretical predictions for the
corrections to the K\"ahler potential presented in \cite{cn} are
given by:
\begin{eqnarray}
\Delta K=-\frac{\xi_{\alpha'}}{(T+\ov{T})^{3/2}}-\frac{\xi_{\rm loop}}{(T+\ov{T})^2}
\ ,
\label{xi1xi2}
\end{eqnarray}
where $\xi_{\alpha'}$ is the coefficient of the $\alpha'$-correction
and $\xi_{\rm loop}$ is the coefficient of the string loop correction.
The leading $\alpha'$-corrections were computed
in \cite{becker} and the coefficient $\xi_{\alpha'}$ was found to be
of the following form:
\begin{eqnarray}
\xi_{\alpha'}=-\frac{\chi\zeta(3)e^{-3\phi_0/2}}{2} \ ,
\end{eqnarray}
where $\chi$ is the Euler number of the compactification manifold
and $\phi_0$ is the expectation value of the dilaton
(which we assume to be stabilized by fluxes). The form of the
leading string loop correction was found in \cite{gersdorff}
by a dimensional analysis.
Further studies of the string loop corrections were done
e.g.\ in \cite{sloop1}-\cite{sloop3} but quite
little is known about the coefficient $\xi_{\rm loop}$.

For those specific corrections (\ref{xi1xi2}), the scalar curvature
$R_T$ reads:
\begin{eqnarray}
R_T=\frac{2}{3}-\frac{35}{48}\frac{\xi_{\alpha'}}{(T+\ov{T})^{3/2}}
-\frac{8}{3}\frac{\xi_{\rm loop}}{(T+\ov{T})^2}+\ldots \ ,
\end{eqnarray}
where ellipsis stands for the higher order terms. The numerical
coefficient in front of the $\alpha'$-correction is comparable with
the leading order contribution, $\frac{2}{3}$, while the numerical
coefficient in front of the string loop correction is four times bigger.
Therefore, there is a chance to find saddle points with positive
trace of the $\eta$-matrix in the region where the corrections
to the K\"ahler potential are small enough to trust the perturbative
expansion, especially with the help of the string loop corrections.

Corrections to the leading order K\"ahler potential have already been
used to implement inflation in type IIB string theory. In \cite{westphal}
a racetrack inflation model was presented in which the
$\alpha'$-corrections were used
to uplift a SUSY AdS minimum to a dS space (and to break supersymmetry).
In the next sections we investigate models in which
the corrections to the K\"ahler potential are used to
implement inflation, but in a different way than in \cite{westphal}.
We start with a SUSY Minkowski (or near-Minkowski)
minimum and then add the K\"ahler
corrections to modify the structure of the $\eta$-matrix in order
to satisfy the slow-roll conditions.
The K\"ahler corrections do not affect the position
of the SUSY Minkowski minimum (see (\ref{minkcond})) so the gravitino
remains massless (or at least very light).

\section{Problems with inflation in Kallosh-Linde model}

It was pointed out in \cite{kl} that in the KKLT-type models,
there is the following relation between the scale of inflation
and the gravitino mass:
\begin{eqnarray}
        \label{gravmass}
        H\leq m_{3/2} \ .
\end{eqnarray}
A typical scale of inflation is much above the TeV scale,
so the above relation makes a low-energy supersymmetry breaking
problematic. This problem is caused by uplifting a AdS minimum to
a Minkowski (or dS) space. In models with a SUSY Minkowski minimum
no uplifting is needed, so this problem is evaded.
The simplest model of this type is the KL model.

In this section we examine the possibility of inflation in the KL model.
The superpotential in this model reads:
\begin{eqnarray}
W=A+Ce^{-cT}+De^{-dT} \ .
\label{racetrack_W}
\end{eqnarray}
The exponential terms come from gaugino condensation in the hidden sector.
The parameters $c=\frac{2\pi}{N}$ and $d=\frac{2\pi}{M}$ are determined
by the rank of the hidden sector gauge group $SU(N)\times SU(M)$.
Without loosing generality, we choose $c>d$.
The parameter $A$ is the contribution from fluxes.
Conditions (\ref{minkcond}) can be solved exactly and imply
a SUSY Minkowski minimum at the following value of $T$:
\begin{eqnarray}
T_{\rm mink}=t_{\rm mink}=\frac{1}{c-d}\ln\left|{\frac{cC}{dD}}\right|
.
\end{eqnarray}
Notice that $\tau_{\rm mink}=0$ at this minimum.
The existence of a SUSY Minkowski minimum imposes the following
constraint on the superpotential parameters:
\begin{eqnarray}
A=-C\left|\frac{cC}{dD}\right|^{\frac{c}{d-c}}
-D\left|\frac{cC}{dD}\right|^{\frac{d}{d-c}}
,
\end{eqnarray}
which we will use to eliminate $A$ from all formulae in this section.
The scalar potential can be written in the form:
\begin{eqnarray}
\label{potential}
6{\cal N}t^2V\br=\br\left[c\left(dt+3\right) {e^{-d\Delta}}
-d \left( 3+ct \right) {e^{-c\Delta}}
-3\left(c-d\right) \right]  \left( {e^{-d\Delta}}-{e^{-c\Delta}}\right)
\nn[4pt]
\br+\br6 \left( c-d \right)
\left( {e^{-d\Delta}} \sin^2 \left( \frac{d\tau}{2} \right)
-{e^{-c\Delta}} \sin^2 \left( \frac{c\tau}{2} \right)  \right)
\nn[4pt]
\br+\br2{e^{- \left( c+d \right)\Delta }} \left( 2cdt+3c+3d
 \right)  \sin^2 \left(\frac{\left( c-d \right)\tau}{2}  \right)
,
\end{eqnarray}
where
${\cal  N}\equiv\left|\frac{cC}{dD}\right|^{\frac{c+d}{c-d}}
\left|CD\right|^{-1}$ and $\Delta\equiv t-t_{\rm mink}$.
The SUSY Minkowski minimum is located at $\Delta=0$, $\tau=0$.
The explicite dependence on the parameters $C$ and $D$
factorizes and appears only in the combination ${\cal N}$.
However, there is a hidden dependence on these parameters in $\Delta$
because $t_{\rm mink}$ depends on the ratio $\frac{C}{D}$.

For the potential (\ref{potential}) one could imagine two scenarios of
inflation starting at a saddle point and ending in the SUSY Minkowski
minimum: one dominated by $t$ field and another dominated by $\tau$.
However, there are obstacles which do not allow for any of these scenarios
to be realized in the KL model.
First of all, as we have shown in the previous section,
for any superpotential, inflation is impossible when the K\"ahler
potential has the standard form given in eq.\ (\ref{kahler}).
But even for a more general K\"ahler potential no slow-roll inflation
can be realized with the racetrack superpotential (\ref{racetrack_W}).
It is instructive to discuss in some detail the KL model with the
uncorrected K\"ahler potential. Then, it will be easier to understand
why for the corrected K\"ahler potential inflation is still not possible.

\subsection{{\boldmath $t$} as candidate for inflaton}

We begin with the case for which $t$ is a candidate for the inflaton.
It is easy to see that for $\tau=0$ the potential (\ref{potential})
has vanishing derivatives: $\left.\frac{\pa V}{\pa\tau}\right|_{\tau=0}=0$,
$\left.\frac{\pa^2 V}{\pa t\pa\tau}\right|_{\tau=0}=0$.
So, at each $t$ for which
$\left.\frac{\partial V}{\partial t}\right|_{\tau=0}=0$,
there is a stationary point
with a diagonal matrix of the second derivatives of the potential.
Some of such points may be saddle points with instability in the $t$
direction.
In order to study such stationary points
we compute the first derivative of (\ref{potential}) with $\tau$
set to zero\footnote{
We choose $\tau=0$ but one should remember that the
potential (\ref{potential}) is periodic in $\tau$ with the period
equal to the smallest common multiple of integers $M$ and $N$
introduced after eq.\ (\ref{racetrack_W}).
}:
\begin{eqnarray}
6{\cal N}t^3\frac{\pa V}{\pa t}
=\left[(ct+2)e^{-c\Delta}-(dt+2)e^{-d\Delta}\right]
\left[c(2dt+3)e^{-d\Delta}-d(2ct+3)e^{-c\Delta}-3(c-d)\right] \ .
\label{V'tau0}
\end{eqnarray}
\begin{figure}[t!]
  \centering
  \includegraphics[width=9cm,angle=0]{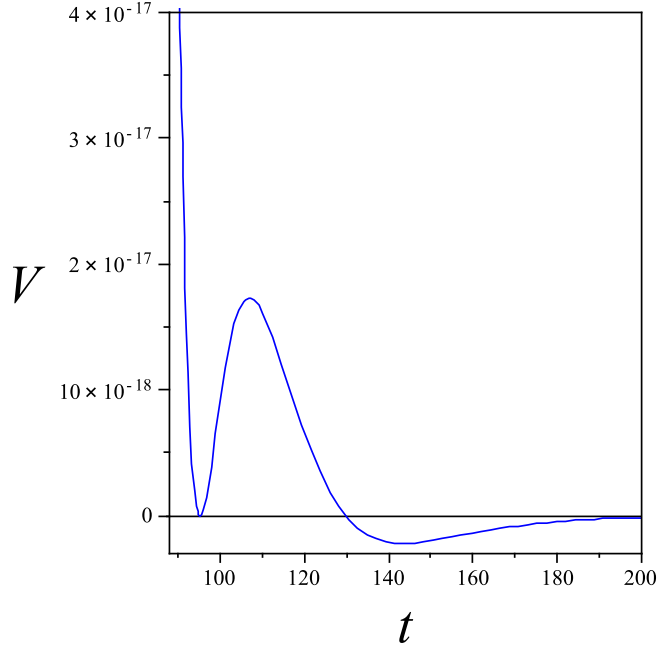}
  \caption{Typical structure of the scalar potential (\ref{potential})
for $\tau=0$.}
  \label{plot1}
\end{figure}
The above function has three zeros which correspond to three extrema
of the potential (see fig.\ \ref{plot1}). The first one is the SUSY
Minkowski minimum ($\Delta=0$) corresponding to the first zero of the
function in the second square bracket in (\ref{V'tau0}).
The second one is a maximum (saddle point from the two-dimensional point
of view) corresponding to the only zero of the function in the first
square bracket. The third one is an AdS minimum which corresponds to the
second zero of the second square bracket.

In principle, there could be a chance that inflation starts at the
vicinity of the saddle point from which the inflaton slowly rolls down
towards the SUSY Minkowski minimum. Therefore, we study this saddle
point in more detail. Its position is given
by the solution of the following equation:
\begin{eqnarray}
\label{pmax}
(ct+2)e^{-c\Delta}-(dt+2)e^{-d\Delta}=0 \ .
\end{eqnarray}
This equation cannot be solved exactly. However, it can be solved
in the limit $ct,dt\gg 1$. The approximate solution is given by:
\begin{eqnarray}
\label{apdel}
\Delta_{\rm sp}\approx\frac{1}{c-d}\ln{\frac{c}{d}} \ .
\end{eqnarray}
For $c>d$ it satisfies the condition $d\Delta_{\rm sp}<1$, while
in the limit $c\rightarrow d$, we obtain $d\Delta_{\rm sp}=1$.
It can be shown that the exact value of $\Delta_{\rm sp}$ is
smaller than its approximate value (\ref{apdel}).

In order to check whether slow-roll inflation is possible,
we compute the parameter $\eta$ at the saddle point (\ref{pmax}).
Since $\eta_t^{\tau}$ vanishes at $\tau=0$, $\eta$-matrix is
diagonal and parameter $\eta=\eta_t^t=\frac{2t^2}{3}\frac{V_{tt}}{V}$
is found to be:
\begin{eqnarray}
\label{etat}
\eta=-\frac{2\left[cdt^2+2(c+d)t+2\right]}{3}
\left(1+\frac{cdt^2e^{-d\Delta_{\rm sp}}}
{\left[cdt^2+3(c+d)t+6\right]e^{-d\Delta_{\rm sp}}-3ct-6}\right) \ .
\end{eqnarray}
One could hope to get $|\eta|\ll1$ by tuning the second
bracket in the above equation to a very small value.
This would require:
\begin{eqnarray}
\label{etatcond}
e^{d\Delta_{\rm sp}}
\approx
1+\frac{dt(2ct+3)}{3ct+6}
\end{eqnarray}
We will show that the above approximate equality can not be fulfilled.
Let us start with the limit $ c\to d$ in which
eq.\ (\ref{pmax}) can be solved for arbitrary value of $t_{\rm mink}$
giving:
\begin{eqnarray}
d\Delta_{\rm sp}\approx\frac{\sqrt{d^2t_{\rm mink}^2
+6dt_{\rm mink}+1}-1-dt_{\rm mink}}{2}
\,.
\end{eqnarray}
This is a monotonically growing function of $dt_{\rm mink}$
with the upper limit $d\Delta_{\rm sp}<1$.
For the minimal possible value $dt_{\rm mink}=1$,
the above equation yields $d\Delta_{\rm sp}=\sqrt{2}-1$.
Therefore, for $dt_{\rm mink}=1$, the l.h.s.\ of (\ref{etatcond})
equals $e^{d\Delta_{\rm sp}}=e^{\sqrt{2}-1}\approx1.5$ and is smaller
than the r.h.s.\ which equals $1+\frac{dt(2ct+3)}{3ct+6}\approx1.8$.
Observing that the r.h.s.\ of (\ref{etatcond}) grows faster with
$dt_{\rm mink}$ than the l.h.s.\ of (\ref{etatcond}), we conclude
that, in the limit $c \to d$, condition (\ref{etatcond}) cannot
be satisfied for any value of $dt_{\rm mink}\ge1$. It is easy to see that
this conclusion remains true also for $c>d$. For fixed $dt$,
the r.h.s.\ of (\ref{etatcond}) increases with growing $ct$ while
$d\Delta_{\rm sp}$, and hence the l.h.s.\ of (\ref{etatcond})
decreases. We have shown that the square bracket in (\ref{etat})
can not be very small.

The range of possible values of the $\eta$ parameter
may be, in some approximation, found from eq.\ (\ref{etat}).
In the limit $ct,dt\gg 1$, we get:
\begin{eqnarray}
|\eta|\approx\frac{4cdt^2}{3}\gg1 \ .
\end{eqnarray}
The smallest possible value of
$|\eta|\approx19$
is obtained for $dt_{\rm mink}=1$ and $c\to d$.
Therefore, $|\eta|$ is never small enough and slow-roll inflation
is not possible.

This result can be interpreted as a manifestation of the well-known
$\eta$-problem in supergravity, which states that a generic value of
$\eta$ is of order one. This problem can be evaded for fields
which do not appear in the K\"ahler potential. This is the case
for the field $\tau$. We examine the possibility that $\tau$ is
an inflaton in the next subsection.

\subsection{{\boldmath $\tau$} as candidate for inflaton}

\begin{figure}[b!]
  \centering
  \includegraphics[width=8.5cm,angle=0]{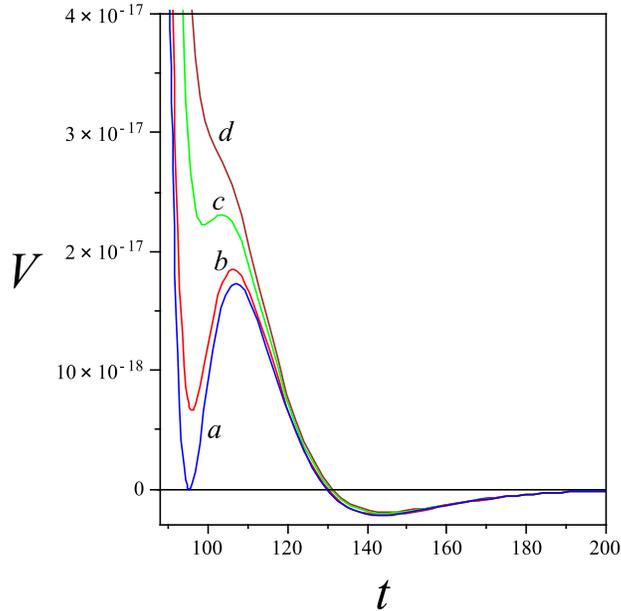}
  \caption{Plot presents disappearing of the barrier for increasing
$\tau$. Different lines correspond to different values
of $c\tau$: (a) $c\tau=0$, (b) $c\tau=0.2$,
(c) $c\tau=0.4$, (d) $c\tau=0.5$.}
  \label{plot2}
\end{figure}
\begin{figure}[t!]
  \centering
  \includegraphics[width=8cm,angle=0]{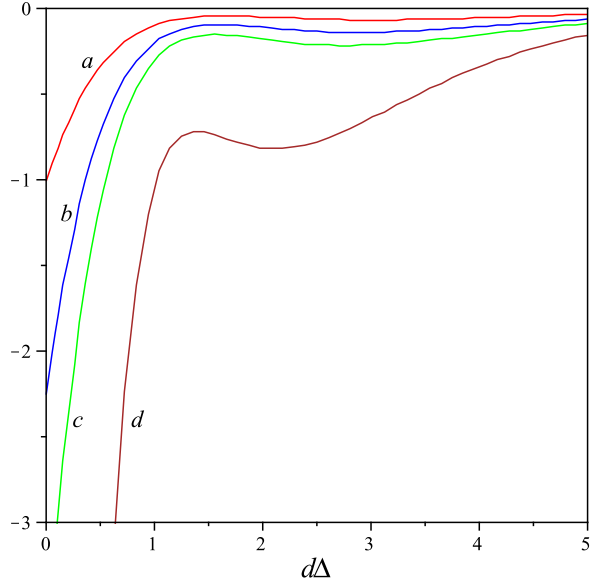}
  \caption{Plot of the expression (\ref{vttau}) for $dt_{\rm mink}=1$.
Different lines correspond to different values of $\delta$:
(a) $\delta=0.1$, (b) $\delta=0.2$, (c) $\delta=0.3$,
(d) $\delta=1$.
}
  \label{plot3}
\end{figure}
The position of the SUSY Minkowski minimum is at $\tau=0$.
Therefore, a saddle point with instability in the $\tau$ direction
has to be situated at $\tau\neq0$. The potential (\ref{potential})
depends on $\tau$ through sines of $c\tau$, $d\tau$ and $(c-d)\tau$,
which have the first maximum in the $\tau$ direction at
$\frac{\pi}{c}$, $\frac{\pi}{d}$ and $\frac{\pi}{c-d}$, respectively.
In the limit $ct_{\rm mink}, dt_{\rm mink}\gg1$, the structure of the maxima
in the $\tau$ direction is dominated by the
$\sin\left(\frac{\left(c-d\right)\tau}{2}\right)$ term
(the last term in (\ref{potential})). Furthermore, the coefficients
multiplying $\sin\left(\frac{c\tau}{2}\right)$
and $\sin\left(\frac{d\tau}{2}\right)$ have opposite signs.
All this suggests that the first maximum in the
$\tau$ direction is around $\tau\approx\frac{\pi}{c-d}$.
We checked numerically that for large $ct_{\rm mink}, dt_{\rm mink}$
the first maximum in the
$\tau$ direction is indeed around $\tau\approx\frac{\pi}{c-d}$.
For smaller $ct_{\rm mink}, dt_{\rm mink}$ the first maximum in the
$\tau$ direction appears at least for $\tau>\frac{\pi}{c}$.
To be a saddle point, such a maximum in the $\tau$ direction
must be a minimum in the $t$ direction.
We have found numerically that the minima in the $t$ direction,
having positive energy, exist only for relatively small
values of $\tau$ (see fig.\ \ref{plot2}).
To understand this fact we compute the first derivative of the
potential (\ref{potential}) for $\tau\neq0$:
\begin{eqnarray}
\label{Vtsin1}
        6{\cal N}t^3\frac{\pa V}{\pa t}
\br=\br
\left[(ct+2)e^{-c\Delta}
-(dt+2)e^{-d\Delta }
\right]
\left[c(2dt+3)e^{-d\Delta }-d(2ct+3)e^{-c\Delta }-3(c-d)
\right]
 \nn[4pt]
        \label{Vtsin2}
        \br+\br6(c-d)\left[\sin^2\left(\frac{c\tau}{2}\right)
(ct+2)e^{-c\Delta }
-\sin^2\left(\frac{d\tau}{2}\right)(dt+2)e^{-d\Delta }\right]
 \nn[4pt]
        \label{Vtsin3}
        \br-\br2\sin^2\left(\frac{(c-d)\tau}{2}\right)
e^{-(c+d)\Delta}\left[(ct+2)c(2dt+3)
+(dt+2)d(2ct+3)\right] \ .
\end{eqnarray}
The potential has, for fixed $\tau$, a minimum in the $t$ direction
at $t=t_{\rm min}$ if $\frac{\pa V}{\pa t}$ vanishes at $t_{\rm min}$
and is positive in some interval $(t_{\rm min}, t_{\rm max})$.
There is no such a minimum if $\frac{\pa V}{\pa t}$ is negative
for every  $t<t_{\rm AdS}$\,, where $t_{\rm AdS}$ is the value
of $t$ at the AdS minimum.
We discuss now in more detail the sign of
$\frac{\pa V}{\pa t}$ for non-zero values of $\tau$.
The third term in (\ref{Vtsin1})
is always negative for $\tau\neq0$. The second term in (\ref{Vtsin1})
is positive for small $d\Delta$ but asymptotically for large
$d\Delta$ it is negative too.
Expanding the second term in (\ref{Vtsin1}) to the second order in $\tau$,
we find that it changes sign for $\Delta$ being the solution
of the following equation:
\begin{eqnarray}
        c^2(ct+2)e^{-c\Delta }-d^2(dt+2)e^{-d\Delta }=0 \ ,
\end{eqnarray}
which, in the limit $ct,dt\gg 1$, is given by:
\begin{eqnarray}
        d\Delta\approx \frac{3}{c/d-1}\ln{\frac{c}{d}}<3 \ .
\end{eqnarray}
Therefore, the $\tau$-dependent part of $\frac{\pa V}{\pa t}$,
given by the last two terms in (\ref{Vtsin1}),
could be positive only for $d\Delta\in(0,3)$.
However, it occurs that the $\tau$-dependent part
of $\frac{\pa V}{\pa t}$
is always negative. In order to show this, we expand
$\frac{\pa V}{\pa t}$                                  in $\tau$:
\begin{eqnarray}
\label{expandvt}
6{\cal N}t^3\frac{\pa V}{\pa t}\br=\br
\left[(ct+2)e^{-c\Delta}
-(dt+2)e^{-d\Delta }\left]\right[c(2dt+3)e^{-d\Delta }-d(2ct+3)e^{-c\Delta }
-3(c-d)\right]
\nn[4pt]
\br+\br\frac{(c-d)\tau^2}{2}\left\{-(c-d)e^{-(c+d)\Delta}\left[(ct+2)c(2dt+3)
+(dt+2)d(2ct+3)\right]
\right.\nn[4pt]
&&
\qquad\qquad
+\left.3\left[c^2(ct+2)e^{-c\Delta }+d^2(dt+2)e^{-d\Delta }\right]\right\}
+{\cal O}(\tau^4)\ .
\end{eqnarray}
We are interested in the coefficient of the $\tau^2$ term.
The negative terms in the square bracket in the above
equation are of order $t_{\rm mink}^2$,
while the positive ones are of order $t_{\rm mink}$
(we recall that the $t_{\rm mink}$-dependence is implicit via
the relation $t=t_{\rm mink}+\Delta$).
Therefore,
for a given $\Delta$, there always exists such $t_{\rm mink}$
that the $\tau$-dependent part of $\frac{\pa V}{\pa t}$ is negative.
Furthermore, the second term of the $\tau$-expansion
(\ref{expandvt}), for $\Delta=0$, is a monotonically
decreasing function of $t_{\rm mink}$:
\begin{eqnarray}
        -\frac{cd(c-d)^2\tau^2t_{\rm mink}(2ct_{\rm mink}
+2dt_{\rm mink}+5)}{2} \ .
\end{eqnarray}
So, if the second term in (\ref{expandvt}) is negative for
the smallest possible value of $dt_{\rm mink}=1$, then it is negative
for any $t_{\rm mink}$. Thus, we concentrate now
on the case $dt_{\rm mink}=1$.
It is convenient to introduce a dimensionless parameter $\delta$:
\begin{eqnarray}
\label{defdelta}
        \delta\equiv\frac{c-d}{d} \ .
\end{eqnarray}
The coefficient of the $\tau^2$ term in (\ref{expandvt}),
up to normalization, can be rewritten in the following way:
\begin{eqnarray}
\label{vttau}
        \br-\br3(3+d\Delta)e^{-d\Delta}
+ 3(1+\delta)^2(d\Delta\delta+\delta+3+d\Delta)e^{-d\Delta(1+\delta) }
\nn[4pt]
        \br-\br
\delta(30+26\delta+5\delta^2+4d^2\Delta^2
+6d^2\Delta^2\delta+2d^2\Delta^2\delta^2+26d\Delta\delta
+7d\Delta \delta^2+22d\Delta)e^{-d\Delta(2+\delta)}
.\quad
\end{eqnarray}
This expression looks quite complicated but after imposing $dt_{\rm mink}=1$,
it is the function of two variables $\delta$ and $\Delta$.
>From fig.\ \ref{plot3} one can see that the second
term in (\ref{expandvt}) is always negative. This is the reason
why for a certain value of $\tau$ the minimum in the $t$ direction disappears.
The potential (\ref{potential}) has no saddle points which are maxima
in the $\tau$ direction.

\subsection{Corrections to K\"ahler potential in KL model}

\begin{figure}[t!]
  \centering
  \includegraphics[width=10cm,angle=0]{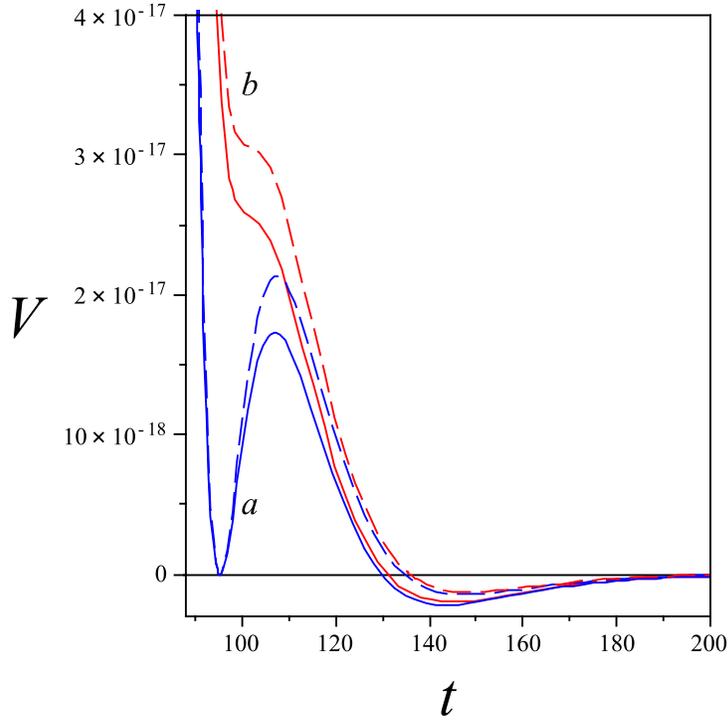}
  \caption{Plot of the potential without corrections (straight lines) and with maximal possible correction $\kappa_{\rm mink}=1$ (dashed lines) for different values of $c\tau$: (a) $c\tau=0$, (b) $c\tau=0.45$.}
  \label{plot8}
\end{figure}

We have shown that the corrections to K\"ahler potential can help in
building models of inflation.
However, as we will show in this subsection,
the corrections to K\"ahler potential are still not sufficient
to implement inflation in the KL model.
For simplicity we use only the $\alpha'$-correction
(the string loop corrections modify the potential in a similar way)
which we denote by $\kappa$:
\begin{equation}
\label{kappa}
\Delta K=-\kappa=-\frac{\xi_{\alpha'}}{(T+\ov{T})^{3/2}} \ .
\end{equation}
The resulting leading correction to the scalar potential reads:
\begin{equation}
\label{delv}
\Delta V =\frac{\kappa}{12(T+\ov{T})^3}
\left|(T+\ov{T})\pa_T W+3W\right|^2 \ ,
\end{equation}
As in the case without corrections, we are most interested
in the $\tau^2$ term of the
expansion\footnote{
The explicite expressions for the potential
and its derivative are given in the appendix.
}
of $\frac{\pa V}{\pa t}$:
\begin{eqnarray}
\label{expandvtcor}
192 \br {\cal N}\br cdt^4\frac{\pa V}{\pa t}=
\nn[4pt]
=\br -\br 16{c}^{2}{
d}^{2} {t}^{3} \left( 4+\kappa \right) \left( {e^{-c\Delta}}c-{e
^{-d\Delta}}d \right)  \left( -{e^{-d\Delta}}+{e^{-c\Delta}} \right)
 \nn[4pt]
\br-\br
4cdt^2 \left\{  \left[ 6\left({c}^2+{d}^{2}\right)
\left( \kappa-4 \right)
+2cd \left(\kappa-32 \right)  \right] {e^{-\Delta \left( d+c \right) }}
-7cd \left( \kappa-8 \right)
 \left( {e^{-2d\Delta}}+{e^{-2c\Delta}} \right)  \right\}
 \nn[4pt]
  \br+\br24cdt^2 \left( c-d \right) \left( \kappa-4 \right)  \left( -{e^{-d
\Delta}}d+{e^{-c\Delta}}c \right)
-81\kappa \left( -d+c+{e^{-c\Delta}}d-{e^{-d\Delta}}c \right) ^{2}
\nn[4pt]
 \br+\br48cdt \left( \kappa-4 \right)  \left( {e^{-d\Delta}}-{e^{-c\Delta}}
 \right)  \left( {e^{-d\Delta}}c-{e^{-c\Delta}}d-c+d \right)
 \nn[4pt]
\br-\br(c-d)cd\tau^2\left\{(c-d)  \left[ 8cd \left( 4+\kappa \right)
\left( d+c \right) {t
}^{3}- \left(  12({c}^{2}+{d}^{2}) \left( \kappa-4 \right)
+4cd \left( \kappa-32 \right)  \right) {t}^{2
}
\right.\right.
\nn[4pt]
&&\qquad\qquad\left.
-24 \left( \kappa-4 \right)  \left( d+c \right) t+81
\kappa \right]\cdot{e^{-\Delta \left( d+c \right) }}
+3 c  \left[ 4ct \left( \kappa-4 \right)( ct+2)
-27\kappa \right]{e^{-c\Delta}}
\nn[4pt]
&&\qquad\qquad\left.
-3d  \left[ 4dt \left( \kappa-4 \right)( dt+2)
-27\kappa \right] {e^{-d\Delta}}\right\} +{\cal O}(\tau^4)  \ ,
\end{eqnarray}
It is a complicated expression but one can figure out some interesting
features. For $\tau=0$, the leading term in the limit $ct,dt\gg1$
(the first term in (\ref{expandvtcor})) factorizes in the same way
as in the uncorrected case (the first term in (\ref{expandvt})).
Therefore, the correction affects
the leading term in $\frac{\pa V}{\pa t}$ only by a small
change of an overall coefficient ($|\kappa|<1$ in
the perturbative regime).
The position of the maximum in the $t$ direction of that leading term
remains unchanged.
Obviously, the corrections to the non-leading terms in (\ref{expandvt})
may change this position a little bit, but in all the terms
the corrections change the coefficients only by small fractions.
Figure \ref{plot8} shows that the position of the barrier is
almost unchanged by the corrections. One can see also that the
corrections make the barrier slightly higher. This decreases
the parameter $\eta$ but only by a small amount. Therefore, we
expect the parameter $\eta$ to be only marginally changed by the
corrections. We confirmed these expectations by numerical analysis.
We conclude that inflation from the saddle point in the $t$ direction
is not possible in the KL model.
\begin{figure}[t!]
  \centering
  \includegraphics[width=9cm,angle=0]{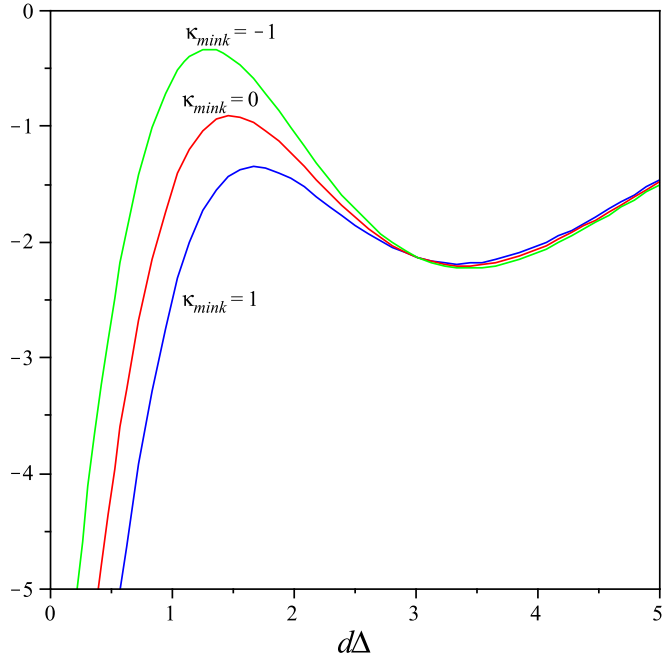}
  \caption{Plot of the $\tau$-dependent part of $\frac{\pa V}{\pa t}$
expanded to the order of $\tau^2$ with the condition $dt_{\rm mink}=1$
imposed.
The value of $\delta=0.2$ is used but the plot does not differ
qualitatively for other values of $\delta$.
$\kappa_{\rm mink}$ is the value of $\kappa$ (defined in (\ref{kappa}))
at the Minkowski minimum. $\kappa_{\rm mink}=0$ corresponds to
the case without
corrections, while $\kappa_{\rm mink}=1$ and $\kappa_{\rm mink}=-1$
correspond to the case of
the correction at the border of validity of the perturbative
expansion. We recall that only corrections with positive
$\kappa$ give positive contribution to the trace of the $\eta$-matrix.}
  \label{plot4}
\end{figure}

The $\tau$-dependent part of $\frac{\pa V}{\pa t}$ is less complicated.
Similarly to the uncorrected case, it can be shown that it is always
negative. We proceed in the same way as before.
For $\Delta=0$, the term proportional
to $\tau^2$ in (\ref{expandvtcor}) is still a monotonically
decreasing function of $t_{\rm mink}$:
\begin{eqnarray}
-8 \left[ (c+d)t_{\rm mink} \left( 4+\kappa \right)+10+\kappa \right]
t_{\rm mink}^{2} \left( d-c \right) ^{2}{c}^{2}{d}^{2}{\tau}^{2} \ .
\end{eqnarray}
So, again it is enough to concentrate on the case
with $dt_{\rm mink}=1$. The term proportional to $\tau^2$ in the
expansion of $\frac{\pa V}{\pa t}$ is given by (\ref{deltakappa})
in the appendix. The plot of this function is presented on fig.\
\ref{plot4}. One can see that even relatively
large corrections do not significantly alter the results as compared
to the uncorrected case. The $\tau$-dependent part of
$\frac{\pa V}{\pa t}$ is still negative.
The minimum in the $t$ direction disappears for values of $\tau$
similar to those in the uncorrected case, as seen in fig.\ \ref{plot8}.

In this section we have shown that it is not possible to implement
inflation in the KL model with or without corrections to the
K\"ahler potential. We conclude that the realization of inflation
requires not only the corrections to the K\"ahler potential but also
some change of the form of the superpotential.

\section{Triple gaugino condensation model}

In this section, we show that  a successful inflation
in the vicinity of a SUSY Minkowski minimum
can be achieved by changing the superpotential.
We consider models with the superpotential containing three
gaugino condensation terms:
\begin{eqnarray}
W=A+Be^{-bT}+Ce^{-cT}+De^{-dT}
\,.
\end{eqnarray}
We allow $A=A_0+i\alpha_0$ and $B=B_0+i\beta_0$ to be complex
(the reasons for such choice will be explained later)
but, for simplicity, assume that $C$ and $D$ are real.
The hidden sector gauge group is $SU(N)\times SU(M)\times SU(L)$
and the additional parameter in the exponential, as compared to
racetrack superpotential, is $b=\frac{2\pi}{L}$.
We take the K\"ahler potential with the leading corrections:
\begin{eqnarray}
K=-3\ln(T+\ov{T})-\frac{\xi_{\alpha'}}{(T+\ov{T})^{3/2}}
-\frac{\xi_{\rm loop}}{(T+\ov{T})^2} \ .
\label{Kcorr}
\end{eqnarray}

In this model
the conditions (\ref{minkcond}) for the existence of a SUSY Minkowski
minimum cannot be solved analytically and the solution is not unique.
A new feature is the possibility of having a SUSY Minkowski minimum
for a non-zero value of $\tau$ (this requires a non-zero
imaginary part of $A$). Moreover, near this SUSY Minkowski minimum
there exists a dS saddle point at which inflation could start.
However, one can check that it is impossible to have a small
$\eta$-parameter if all three parameters $B$, $C$ and $D$ are real.
We performed numerical analysis of the $\eta$-matrix for many different
sets of the parameters: $B$, $C$, $D$, $b$, $c$, $d$, $\xi_{\alpha'}$ and
$\xi_{\rm loop}$, adjusting parameter $A$ to keep $W=0$ at the Minkowski minimum.
We observed that the behaviour of the $\eta$-matrix is similar
if we change any of the parameters $B$, $C$, $D$, $b$, $c$ or $d$.
For concreteness, let us concentrate on changing parameter $B$
keeping other (except $A$) fixed. For a certain value of $B$,
the SUSY Minkowski minimum is situated at $\tau=0$ (as in the KL model)
and the saddle point is unstable in the $t$ direction having very large,
negative $\eta_t^t$. Changing $B$, we can move the SUSY Minkowski
minimum to a non-zero value of $\tau$, but a nearby saddle point
has very large negative $\eta_{\tau}^{\tau}$. Changing $B$ further,
one can obtain very small negative $\eta_{\tau}^{\tau}$.
Unfortunately, the off-diagonal $\eta_t^{\tau}$ entry is very large,
so the parameter $\eta$ remains also very large. For a small range
of $B$ both diagonal entries of the $\eta$-matrix are positive but
the trace is always smaller than the off-diagonal entry. This implies
again a large, negative $\eta$-parameter. Changing $B$ one can
also obtain very small, negative $\eta_t^t$ but still large
off-diagonal entry prevents from the slow-roll regime.
The parameters $\xi_{\alpha'}$ and $\xi_{\rm loop}$, which parametrize
the corrections to the K\"ahler potential, change the $\eta$-matrix
in a different way. All the entries of the $\eta$-matrix grow with
increasing $\xi_{\alpha'}$ or $\xi_{\rm loop}$. The trace
of the $\eta$-matrix also
grows but still it is smaller than the off-diagonal entry.
We conclude that the main obstacle in obtaining a flat saddle
point are large off-diagonal entries of the $\eta$-matrix.

The situation changes when one allows for a small imaginary parts
of $B$, $C$ or $D$. We choose $B=B_0+i\beta_0$ to be complex but
similar results can be obtained when choosing $C$ or $D$ to be
complex\footnote{
Nonzero imaginary parts of $B$, $C$ and $D$ can originate from loop
threshold corrections to the corresponding gauge kinetic functions
(see e.g. \cite{gaugino_corr1}-\cite{gaugino_corr2}).
Such corrections usually depend logarithmically on various moduli
fields and can modify gaugino condensation terms by moduli-dependent
polynomial prefactors. In our model most of the moduli are assumed
to be stabilized at higher scales. If some of them has complex vevs,
the parameters $B$, $C$, $D$ can have imaginary parts. Moreover,
relative phases of $B$, $C$ and $D$ can be changed by redefining
the axion $\tau$.
}.
Changing the value of $\beta_0$, one can obtain a very small
off-diagonal entry of the $\eta$-matrix. Furthermore,
with one more parameter ($B_0\,$, $C$ or $D$) fine-tuned,
a slow-roll inflation is possible.
For the numerical example we choose the following set of parameters:
\begin{eqnarray}
\label{par}
B_0=-5.454364\cdot10^{-2}\ ,\hspace{1cm}\beta_0=5.939476\cdot10^{-5}
\ ,\hspace{1cm}C=-\frac{1}{75}\ ,\hspace{1cm}D=\frac{1}{30}\ ,
\nn[4pt]
b=\frac{2\pi}{70}\ ,\hspace{1cm}c=\frac{2\pi}{100}
\ ,\hspace{1cm}d=\frac{2\pi}{90}\ ,\hspace{1cm}\xi_{\alpha'}=500
\ ,\hspace{1cm}\xi_{\rm loop}=5000\ .
\end{eqnarray}

For this set of parameters the condition $W\approx0$ at the SUSY
near-Minkowski minimum is obtained by tuning $A_0$ and $\alpha_0$
in the following way:
\begin{eqnarray}
\label{par2}
A_0=7.2058574\cdot10^{-7}\ ,\hspace{2cm}\alpha_0=-9.4134768\cdot10^{-8} \ .
\end{eqnarray}
The exact SUSY Minkowski minimum can be obtained
only by exact tuning $W=0$. However, in our world SUSY is broken
and the gravitino mass does not vanish. Therefore, tuning of $A$
does not have to be very precise.

In this example we use both $\alpha'$ and string loop corrections but
an inflationary saddle point can be found using $\alpha'$ or string loop
correction alone. Therefore, our effective model is also valid
for Calabi-Yau compactifications for which one of these corrections
is suppressed.

\begin{figure}[t!]
  \centering
  \includegraphics[width=9.5cm,angle=0]{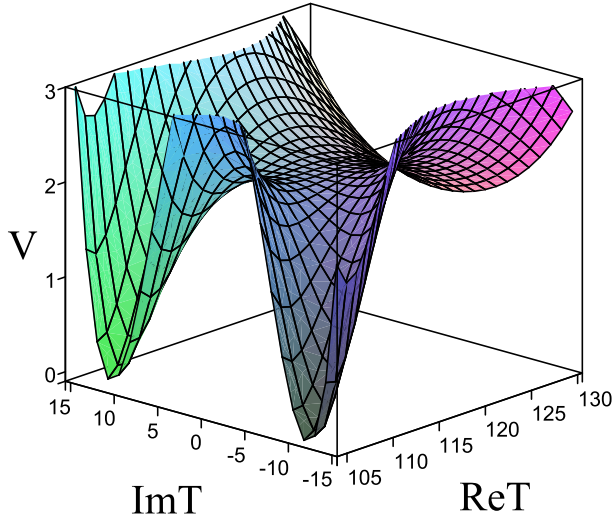}
  \caption{Plot of the inflationary part of the
potential multiplied by $10^{19}$ in the triple
gaugino condensation model for parameters (\ref{par})-(\ref{par2}).}
  \label{plot5}
\end{figure}

The structure of the inflationary potential is shown at fig.\ \ref{plot5}.
There are two minima and the saddle point where inflation can take place.
One of these minima is of AdS-type and is situated at the following
field values:
\begin{eqnarray}
t_{\rm AdS}=104.646\ , \hspace{2cm} \tau_{\rm AdS}=-11.664 \ .
\end{eqnarray}
We are more interested in the SUSY Minkowski minimum which occurs at:
\begin{eqnarray}
\label{minknum}
t_{\rm mink}=104.473\ , \hspace{2cm} \tau_{\rm mink}=10.885 \ .
\end{eqnarray}
The inflationary saddle point is situated at:
\begin{eqnarray}
\label{sadnum}
t_{\rm saddle}=115.475\ , \hspace{2cm} \tau_{\rm saddle}=-0.183 \ .
\end{eqnarray}
In order to trust the perturbative expansion of the K\"ahler
potential, the ratios ${\xi_{\alpha'}}/{(T+\ov{T})^{3/2}}$ and
${\xi_{\rm loop}}/{(T+\ov{T})^2}$ has to be small. For the
Minkowski minimum (\ref{minknum}) these turn out to be,
respectively, around $0.17$ and $0.11$, while for the
saddle point (\ref{sadnum}) around $0.14$ and $0.09$.
Therefore, both types of corrections are indeed small.
There are no firm predictions for the values of parameters
$\xi_{\alpha'}$ and $\xi_{\rm loop}$. We have chosen them
in such a way that both give similar corrections to the K\"ahler
potential in the region of the parameter space important for
inflation. Of course, inflation can be realized also for other
values of those parameters, e.g. for slightly bigger
$\xi_{\alpha'}$ and much smaller $\xi_{\rm loop}$
(for large values of $t$, the coefficient multiplying
$\xi_{\alpha'}$ in eq.\ (\ref{Kcorr}) is much bigger than the
one multiplying $\xi_{\rm loop}$).

To check whether the saddle point (\ref{sadnum}) is
flat enough for inflation, we compute the $\eta$-matrix:
\begin{equation}
\eta\approx\begin{pmatrix}
                1.99688 & 0.097453 \\
        0.097453 & 0.0018433
        \end{pmatrix} \ .
\end{equation}
It has the eigenvalues: $\eta_1= -0.0029058$, $\eta_2=2.00163$.
They differ by three orders of magnitude, so the isocurvature
fluctuations are very small and can be neglected.

\begin{figure}[t!]
  \centering
  \includegraphics[width=13cm,angle=0]{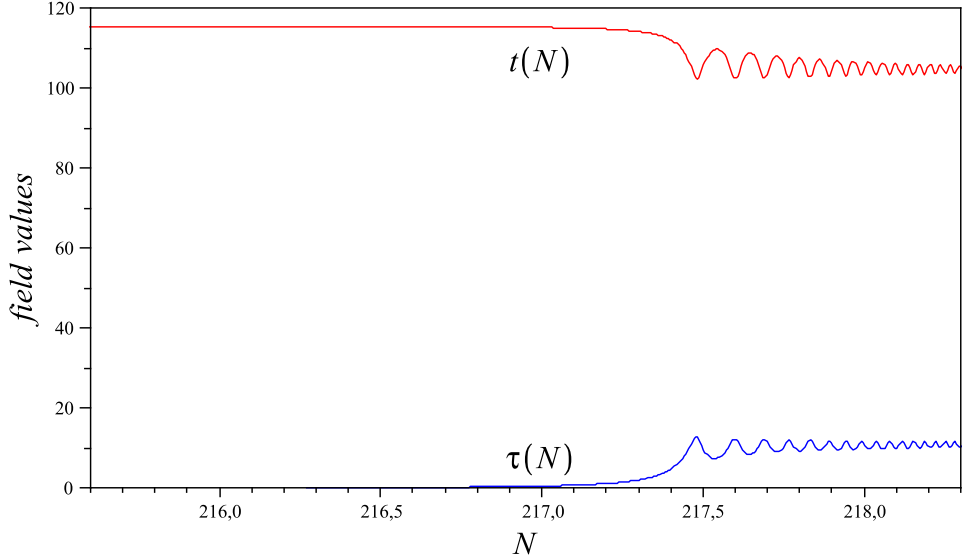}
  \caption{Evolution of the fields in the last stage of inflation
as a function of e-folds $N$.}
  \label{plot6}
\end{figure}

In order to study the evolution of the fields during inflation we have
to solve the appropriate equations of motion which, for non-canonically
normalized fields, are given by:
\begin{eqnarray}
        \ddot{\phi^i}+3H\dot{\phi^i}+\Gamma_{jk}^i\dot{\phi^j}\dot{\phi^k}
+g^{ij}\frac{\pa V}{\pa \phi^i}=0\,, \nn
H=\left(\frac{\dot{a}}{a}\right)^2
=\frac{8\pi G}{3}\left(\frac{1}{2}g_{ij}\dot{\phi^i}\dot{\phi^j}+V\right)\,,
\end{eqnarray}
where $a$ is the scale factor, $H$ is the Hubble parameter and dots
denote derivatives with respect to
cosmic
time.
It is convenient to study the field evolution using the number
of e-folds $N$:
\begin{equation}
        a(t)=e^N, \hspace{2cm}\frac{d}{dt}=H\frac{d}{dN}\,.
\end{equation}
Then, the equations of motion in our model read:
\begin{eqnarray}
t''\br=\br-\left[3-\frac{g_{tt}}{2}\left(t'^2+\tau'^2\right)\right]
\left(g^{tt}\frac{V_t}{V}+t'\right)-\frac{g^{tt}g_{ttt}}{2}
\left(t'^2-\tau'^2\right) \ ,\\
\tau''\br=\br-\left[3-\frac{g_{tt}}{2}\left(t'^2+\tau'^2\right)\right]
\left(g^{tt}\frac{V_{\tau}}{V}+\tau'\right)-g^{tt}g_{ttt}t'\tau' \ ,
\end{eqnarray}
where $g_{tt}=(g^{tt})^{-1}=\frac{1}{2}\frac{\pa^2 K}{\pa t^2}$,
$g_{ttt}=\frac{1}{2}\frac{\pa^3 K}{\pa t^3}$ and $'$ denotes
derivatives with respect to $N$. We solve numerically these equations
for parameters (\ref{par})-(\ref{par2}) with a starting point near
the saddle point (\ref{sadnum}):
\begin{eqnarray}
t(0)=t_{\rm saddle}\ ,\hspace{1cm} \tau(0)=\tau_{\rm saddle}(1-\iota)
\ ,\hspace{1cm}t'(0)=\tau'(0)=0 \ .
\end{eqnarray}
With the initial conditions fine-tuned at the level of $\iota=0.01$,
we obtain about $217$ e-folds of slow-roll inflation before the inflaton
starts to oscillate around the SUSY Minkowski minimum (\ref{minknum}),
as seen in fig.\ \ref{plot6}. For $\iota=0.02$ one can obtain
about $108$ e-folds of inflation which is also enough to explain
flatness and isotropy of the observed Universe.

We should comment on the fine-tuning of this model. First of all,
one should fine-tune $W=0$ to obtain vanishing cosmological constant
and vanishing gravitino mass at the same time. This is done by
tuning the complex parameter $A$. In our world SUSY is broken
and the gravitino mass is nonzero, so this fine-tuning can be
relaxed to some extent. For the low energy SUSY breaking with
$m_{3/2}\approx{\cal O}(1 {\rm TeV})$, the fine-tuning of $A$ is at the
level of $10^{-4}$. For larger gravitino masses the fine-tuning of $A$
is even smaller. On the other hand, for smaller fine-tuning of $A$
we obtain more negative energy in the minimum and stronger uplifting
is needed to obtain a vanishing (or slightly positive) cosmological
constant. We recall that uplifting of a deep AdS minimum is the
source of large gravitino masses in typical KKLT-type models.
In our model the AdS minimum is not very deep. We studied numerically the
effect of uplifting on the gravitino mass. We used effective uplifting term
$\Delta V=\frac{E}{t^2}$ and found that such uplifting changes the
gravitino mass by extremely small amount. For example
for $m_{3/2}\approx{\cal O}(1 {\rm TeV})$ the uplifting
changes the gravitino mass only by around $10^{-19}$.

Secondly, one has to fine-tune $B_0$ and $\beta_0$
to ensure the flatness of the inflationary saddle point.
$B_0$ is fine-tuned at the level of $10^{-7}$, while
fine-tuning of $\beta_0$ is at the level of $10^{-5}$.
In the original racetrack model \cite{racetrack} there is only one
fine-tuning at the level of $10^{-4}$ needed to obtain small
$\eta$-parameter. Therefore, we conclude that,
at least in our triple gaugino condensation model,
the price for small gravitino mass is an additional
fine-tuning of parameters. It would be very interesting to check
whether this additional fine-tuning is a general feature of the
KKLT-type models with a small gravitino mass.
We leave it for the future work.

\subsection{Rescaling properties}

The triple gaugino condensation model has some rescaling properties.
There are some transformations of the parameters that do not affect
the potential or scale the potential in such a way that
the slow-roll parameters remain unchanged. One of them reads:
\begin{eqnarray}
\label{scale1}
        A\rightarrow kA
\ ,\hspace{1cm} B\rightarrow kB
\ ,\hspace{1cm} C\rightarrow kC\ ,\hspace{1cm} D\rightarrow kD
\end{eqnarray}
with other parameters and field $T$ unchanged.
This transformation scales the potential and the amplitude of the
density perturbations $\frac{\delta\rho}{\rho}$ in the following way:
\begin{eqnarray}
\label{scale1a}
V\rightarrow k^2 V\ , \hspace{1cm}
\frac{\delta\rho}{\rho}\rightarrow k\frac{\delta\rho}{\rho}\ .
\end{eqnarray}
Another transformation is given by:
\begin{eqnarray}
\label{scale2}
b\rightarrow\lambda^{-1} b\ ,\hspace{1cm} c\rightarrow\lambda^{-1} c
\ ,\hspace{1cm} d\rightarrow\lambda^{-1} d
\ ,\hspace{1cm} \xi_{\alpha'}\rightarrow\lambda^{3/2} \xi_{\alpha'}
\ ,     \hspace{1cm} \xi_{\rm loop}\rightarrow\lambda^{2} \xi_{\rm loop}
\end{eqnarray}
with other parameters unchanged. If field T is also rescaled:
\begin{eqnarray}
T\rightarrow\lambda T
 \ ,
\end{eqnarray}
then, the potential and the amplitude of density perturbations
scale as:
\begin{eqnarray}
\label{scale2a}
        V\rightarrow\lambda^{-3} V
\ , \hspace{1cm}
\frac{\delta\rho}{\rho}\rightarrow\lambda^{-3/2}\frac{\delta\rho}{\rho}
 \ .
\end{eqnarray}
The above two transformations can be used to change the parameters
in order to have a correct amplitude of density perturbation.
Combining these transformations one can obtain a transformation that
do not change the potential at all:
\begin{eqnarray}
\label{scale3}
        A\br\rightarrow\br \zeta A
\ ,\hspace{1cm} B\rightarrow \zeta B
\ ,\hspace{1cm} C\rightarrow \zeta C\ ,\hspace{1cm} D\rightarrow \zeta D\ ,
\nn[4pt]
        b\br\rightarrow\br\zeta^{-2/3} b
\ ,\hspace{1cm} c\rightarrow\zeta^{-2/3} c
\ ,\hspace{1cm} d\rightarrow\zeta^{-2/3} d
\ ,\nn[4pt]
 \xi_{\alpha'}\br\rightarrow\br\zeta \xi_{\alpha'}
\ ,     \hspace{1cm} \xi_{\rm loop}\rightarrow\zeta^{4/3} \xi_{\rm loop}\ ,
\hspace{1cm}
 T\rightarrow\zeta^{2/3} T \ .
\end{eqnarray}
This transformation is very useful because it does not change
any predictions of the model. If one of the parameters chosen in
our example do not fulfill string-theoretical constraints
(which hopefully will appear in the near future) one will be able
to use (\ref{scale3}) to change this parameter accordingly
without changing inflationary predictions.
In particular, with the help of this transformation, one can reduce
the rank of the hidden sector gauge groups, which appear in parameters
$b$, $c$ and $d$.

\subsection{Experimental constraints and signatures}

\begin{figure}[b!]
  \centering
  \includegraphics[width=11.5cm,angle=0]{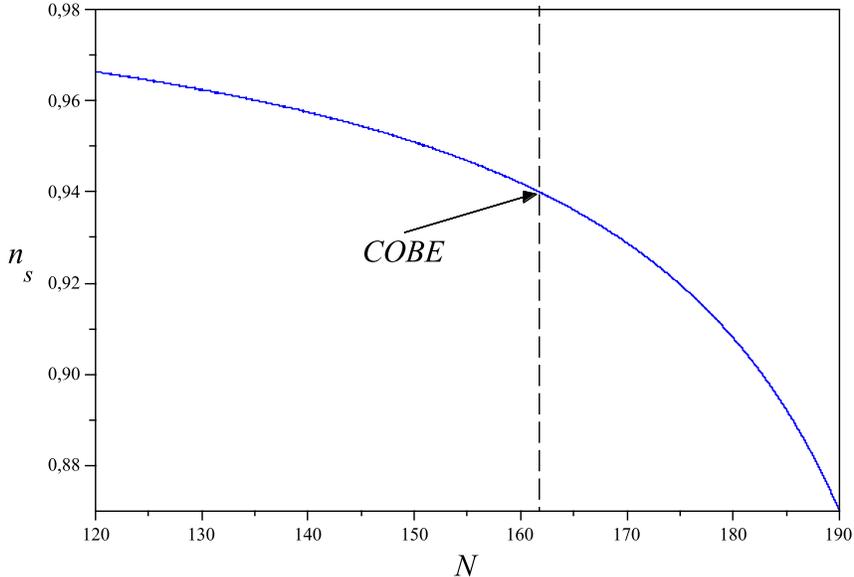}
  \caption{Evolution of the spectral index during inflation
as a function of e-folds $N$ for parameters
(\ref{par})-(\ref{par2}).
The vertical dashed line corresponds to the COBE normalization
point at which we obtain $n_s\approx0.94$.}
  \label{plot7}
\end{figure}

Every inflationary model has to satisfy the COBE normalization \cite{lyth}:
\begin{eqnarray}
\label{cobe}
 \frac{\delta\rho}{\rho}=\frac{2}{5}\sqrt{{\cal P_R}(k_0)}
\approx 2\cdot10^{-5}
\ ,
\end{eqnarray}
where $k_0\approx7.5H_0$
is the scale for which COBE
satellite has measured the amplitude of the density perturbations.
${\cal P_R}$ is the amplitude of the scalar perturbations and is given,
in the slow-roll approximation, by the following formula:
\begin{eqnarray}
{\cal P_R}(k)=\frac{1}{24\pi^2}
\left.\left(\frac{V}{\epsilon}\right)\right|_{k=aH} \ ,
\end{eqnarray}
where the r.h.s.\ of this equation is evaluated at the time when
the scale $k$ crosses the horizon and $\epsilon$ is the generalized
slow-roll parameter, given by:
\begin{eqnarray}
\epsilon\equiv\frac{1}{2}g^{ij}\left(\frac{V_iV_j}{V^2}\right) \ ,
\end{eqnarray}
which in our model reads:
\begin{eqnarray}
\epsilon=\left(\frac{\pa^2 K}{\pa t^2}\right)^{-1}
\left(\frac{V_t^2+V_{\tau}^2}{V^2}\right) \ .
\end{eqnarray}
The COBE normalization has to be imposed approximately $55$ e-folds
before the end of
inflation\footnote{The exact number of e-folds before the end of
inflation for which $k_0$ crosses the horizon depends on the reheating
temperature, which is model-dependent.}.
We have checked that for the parameters (\ref{par})-(\ref{par2}) our
model predicts the amplitude of the density perturbations
about $2.2\cdot10^{-5}$ which is consistent
with COBE measurements (\ref{cobe}).

An important quantity, which gives constraints on inflationary models,
is the spectral index:
\begin{eqnarray}
        n_s-1\equiv\frac{d\ln\mathcal{P}_{\mathcal{R}}(k)}{d\ln k}
\approx\frac{d\ln\mathcal{P}_{\mathcal{R}}(N)}{dN}\ ,
\end{eqnarray}
where the last approximation comes from the fact that this
quantity is evaluated at horizon crossing
$k=aH= He^N$ which implies $d\ln k\approx dN$.
In fig.\ \ref{plot7} we plot $n_s$ versus $N$. The spectral index
at the COBE normalization point has the value:
\begin{eqnarray}
n_s\approx0.94 \ ,
\label{ns}
\end{eqnarray}
which is consistent with the 3-year WMAP result $n_s=0.958\pm0.016$
\cite{wmap}. The value of the spectral index (\ref{ns}) agrees
 also with the results of \cite{ns}, where the bound $n_s\lesssim0.95$
was found for quite general racetrack inflation models
(including models with more than two exponential terms in
the superpotential).
From the slope on fig.\ \ref{plot7} one can see that
$\frac{d n_s}{d\ln k}\approx-0.001$ which is far below the current
upper
limit. The tensor to scalar ratio in the slow-roll approximation
is given by $r=16\epsilon$. In this model we found the value
$r\approx10^{-12}$ which is negligible.

\vspace{-2mm}

\section{Conclusions}

In this paper we have studied the possibility of implementing inflation
ending in a SUSY (near) Minkowski minimum and dominated by one modulus
in type IIB flux compactification.
We have identified a general obstacle that can make a slow-roll
inflation impossible in this type of models. We have found that, for the
tree-level K\"ahler potential, in the one-modulus case, the trace
of the $\eta$-matrix at any non-SUSY stationary point is negative.
In the case of the overall volume modulus this trace is equal
$-\frac{4}{3}$. This implies that the slow-roll parameter $\eta$
near a
non-SUSY saddle point is necessarily smaller than $-\frac{4}{3}$
which is not consistent with the slow-roll condition $|\eta|\ll1$.
Analogous results are valid for other moduli when the K\"ahler potential
has the form given in eq.\,(\ref{standardK}) with $n_X\leq3$.
It is important to stress that this is a general result,
independent of the form of the superpotential.

In order to cure this problem one has to add some corrections to the
K\"ahler potential, such as the $\alpha'$-corrections or the string
loop corrections. However, in the KL model with the
racetrack superpotential inflation cannot be
realized even with a corrected K\"ahler potential. In this model,
Re$T$ is not a good candidate for the inflaton, because it appears
in the K\"ahler potential and therefore suffers from the usual
$\eta$-problem.
Also Im$T$ cannot play the role of the inflaton because saddle points
with instability in the Im$T$ direction do not exist in the KL model.
It turns out that minima in the Re$T$ direction (with positive
value of the potential) disappear
for values of Im$T$ much smaller than that of the
first maximum in the Im$T$ direction.
Therefore, in the class of models
under consideration, not only corrections to the K\"ahler potential
are necessary but one needs also to change the superpotential.

In this paper we have proposed a novel inflationary model with a
triple gaugino condensation. It contains three exponential terms in the
superpotential. The corrections to the K\"ahler potential are
crucial in this model. In the presented example, the
$\alpha'$-corrections and the string loop corrections to
the K\"ahler potential are used
but a successful slow-roll inflation can also be obtained with
only one type of such corrections.
The imaginary part of $T$ plays the role of the inflaton.
More than 100 e-folds can be obtained if the initial
value of Im$T$ is close to the position of the saddle point and
tuned at the level of 0.02.
The spectral index $n_s\approx0.94$ is consistent
with the CMB measurements.

The main distinctive feature of this model is that
the gravitino mass is much smaller than the Hubble
constant during inflation.
However, the price for a small
gravitino mass is an additional fine tuning of the parameters
as compared to the original racetrack inflation \cite{racetrack}.
We leave for a future investigation the question
whether this additional fine-tuning is a generic feature
of inflationary models with a SUSY Minkowski vacuum.
Another interesting subject is how the present analysis
can be generalized to models with more fields.

\section*{Acknowledgments}

The work of M.B.\ was partially supported by
the EU 6th Framework Program MRTN-CT-2004-503369
``Quest for Unification''.
M.O.\ acknowledges partial support from
the EC Project MTKD-CT-2005-029466
``Particle Physics and Cosmology: the Interface''.
and from the Polish MNiSW grant N202 176 31/3844 for years
2006-2008.

\section*{Appendix}
\renewcommand{\theequation}{A.\arabic{equation}}
\setcounter{equation}{0}

In the appendix we give the full expressions for the scalar
potential and its derivative for the modified KL model
with the correction to the K\"ahler potential parametrized by the
parameter $\kappa$ defined in (\ref{kappa}).
The potential has the form:
\begin{eqnarray}
\label{potkappa}
96 {\cal N} cdt^3V\br=\br4{c}^{2}{d}^{2}t^2  \left( 4+\kappa \right)
 \left( {e^{-d\Delta}}-{e^{-c\Delta}} \right) ^{2}
+9\kappa \left( c{e^{-d\Delta}}-d{e^{-c\Delta}}-c+d \right) ^{2}
\nn[4pt]
\br+\br
12\left( \kappa-4 \right) cdt \left( {e^{-d\Delta}}-{e^{-c\Delta}}
 \right)  \left( d{e^{-c\Delta}}-c{e^{-d\Delta}}+c-d \right)
\nn[4pt]
\br+\br4cd{e^{-\Delta \left( c+d \right) }} \left[ 4cd \left( 4+\kappa
 \right) {t}^{2}-6 \left( \kappa-4 \right)  \left( c+d \right) t+9
\kappa \right]  \sin^2 \left( \frac{\left( c-d \right) \tau}{2} \right)
\nn[4pt]
\br+\br12 \left( c-d \right) d{e^{-c\Delta}} \left[ 2ct \left( \kappa-4
 \right) -3\kappa \right]  \sin^2 \left( \frac{c \tau}{2} \right)
\nn[4pt]
\br-\br12 \left( c-d \right) c{e^{-d\Delta}} \left[ 2dt \left( \kappa-4
 \right) -3\kappa \right]  \sin^2 \left( \frac{d \tau}{2} \right) \,.
\end{eqnarray}
Its first derivative with respect to $t$ is given by:
\begin{eqnarray}
\label{dvtalpha'}
192{\cal N}cdt^4\frac{\pa V}{\pa t}\br=\br -16{c}^{2}{
d}^{2} {t}^{3} \left( 4+\kappa \right) \left( {e^{-c\Delta}}c-{e
^{-d\Delta}}d \right)  \left( -{e^{-d\Delta}}+{e^{-c\Delta}} \right)
\nn[4pt]
\br-\br4cdt^2 \left\{  \left[ 6\left({c}^2+{d}^{2}\right)
\left( \kappa-4 \right)
+2cd \left(\kappa-32 \right)  \right] {e^{-
\Delta \left( d+c \right) }}\right.
\nn[4pt]
&&\qquad
-7cd \left.\left( \kappa-8 \right)
 \left( {e^{-2d\Delta}}+{e^{-2c\Delta}} \right)  \right\}
 \nn[4pt]
  \br+\br24cdt^2 \left( c-d \right) \left( \kappa-4 \right)
\left( -{e^{-d
\Delta}}d+{e^{-c\Delta}}c \right)
\nn[4pt]
 \br+\br48cdt \left( \kappa-4 \right)  \left( {e^{-d\Delta}}-{e^{-c\Delta}}
 \right)  \left( {e^{-d\Delta}}c-{e^{-c\Delta}}d-c+d \right)
 \nn[4pt]
\br-\br81\kappa \left( -d+c+{e^{-c\Delta}}d-{e^{-d\Delta}}c \right) ^{2}
\nn[4pt]
\br-\br4cd  \left\{ 8cd \left( 4+\kappa \right)  \left( d+c \right) {t
}^{3}- \left[  12({c}^{2}+{d}^{2}) \left( \kappa-4 \right)
+4cd \left( \kappa-32 \right)  \right] {t}^{2}\right.
\nn[4pt]
&&\quad-24 \left.\left( \kappa-4 \right)  \left( c+d \right) t+81
\kappa \right\}
 {e^{-\Delta \left( d+c \right) }}
\sin^2 \left( \frac{\left( c-d \right) \tau}{2} \right)
 \nn[4pt]
 \br-\br12 \left( c-d \right)d {e^{-c\Delta}}
 \sin^2 \left( \frac{ c\tau}{2}
 \right) \left( 4ct \left( \kappa-4 \right)( ct+2)
-27\kappa \right)
\nn[4pt]
\br+\br12\left( c-d \right)c {e^{-d\Delta}} \sin^2 \left( \frac{ d\tau}{2}
 \right)  \left( 4dt \left( \kappa-4 \right)( dt+2)
-27\kappa \right)   \,.
\end{eqnarray}
The coefficient of the $\tau^2$ term in the expansion of $\frac{\pa V}{\pa t}$ (\ref{expandvtcor}),
after imposing $dt_{\rm mink}=1$ and using parameter $\delta$ defined in (\ref{defdelta}), reads:
\begin{eqnarray}
\label{deltakappa}
\br-\br8{d}^{4}\delta \left( 1+\delta \right)  \left\{ \frac{3}{2}
 \left[  \left( d\Delta+1 \right) ^{2} \left( \kappa-4 \right) {
\delta}^{2}+2 \left( d\Delta+2 \right)  \left( d\Delta+1 \right)
 \left( \kappa-4 \right) \delta
 \right.\right.
 \nn[4pt]
 &&\qquad\qquad\qquad\quad\left.+ \left( \kappa-4
 \right) {d}^{2}{\Delta}^{2}+4 \left( \kappa-4 \right) d\Delta-12-{\frac {15}
{4}}\kappa \right]  \left( 1+\delta \right) {e^{-d\Delta \left( 1+\delta
 \right) }}
 \nn[4pt]
  &&\qquad\qquad\qquad+\left[  \left(  \left( 4+\kappa \right) d\Delta+
10-\frac{1}{2}\kappa \right)  \left( d\Delta+1 \right) ^{2}{\delta}^{2}+2 \left( 4+
\kappa \right) {d}^{3}{\Delta}^{3} \right.
 \nn[4pt]
 &&\qquad\qquad\qquad\quad+3
 \left( d\Delta+1 \right)  \left(  \left( 4+\kappa
 \right) {d}^{2}{\Delta}^{2}+\frac{5}{6} \left( \kappa+{\frac {104}{5}} \right)
d\Delta+{\frac {52}{3}}-\frac{7}{6}\kappa \right) \delta
\nn[4pt]
&&\qquad\qquad\qquad\quad\left.+\frac{5}{2} \left( \kappa+{\frac {104}{5}} \right) {
{d}^{2}\Delta}^{2}+ \left( 104-7\kappa \right) d\Delta+{\frac {21}
{8}}\kappa+60 \right] \delta{e^{- d\Delta \left( 2+\delta
 \right) }}
 \nn[4pt]
 &&\qquad\qquad\qquad\left.-\frac{3}{2} \left[  \left( \kappa-4 \right) {d}^{2}{\Delta}^{2}
+4 \left( \kappa-4 \right) d\Delta-12-{\frac {15}{4}}\kappa
 \right] {e^{-d\Delta}}  \right\} \,.
\end{eqnarray}
Notice that $\kappa$ is not a constant and depends on $\Delta$ too. The plot of the above function is presented on fig.\,\ref{plot4}. One can see that (\ref{deltakappa}) is always negative. This is the reason why for a certain value of $\tau$ the minimum in the $t$ direction disappears and the potential (\ref{potkappa}) has no saddle points which are maxima in the $\tau$ direction.

\end{document}